\documentclass[11pt,preprint]{emulateapj}

\usepackage{hyperref}
\usepackage{amsmath}

\shorttitle{Large-scale and long-term planet migration}
\shortauthors{Ben\'itez-Llambay et al.}

\begin{document}

\title{Long-term and large-scale hydrodynamical simulations of
  migrating planets}

\author{Pablo Ben\'itez-Llambay}
\email{pbllambay@oac.unc.edu.ar} 
\author{Ximena S. Ramos} 
\email{xramos@oac.unc.edu.ar} 

\author{Cristian Beaug\'e} \affil{Instituto de Astronom\'\i a
  Te\'orica y Experimental (IATE), Observatorio Astron\'omico,
  Universidad Nacional de C\'ordoba. Laprida 854, X5000BGR, C\'ordoba,
  Argentina} \email{beauge@oac.unc.edu.ar}

\author{Fr\'ed\'eric S. Masset} \affil{Instituto de Ciencias F\'\i
  sicas, Universidad Nacional Aut\'onoma de M\'exico (UNAM),
  Apdo. Postal 48-3,62251-Cuernavaca, Morelos, M\'exico}
\email{masset@icf.unam.mx}

\begin{abstract}
We present a new method that allows long-term and large-scale
hydrodynamical simulations of migrating planets over a grid-based
Eulerian code.  This technique, which consists in a remapping of the
disk by tracking the planetary migration, enables runs of migrating
planets over a time comparable to the age of protoplanetary disks.
This method also has the potential to address efficiently problems
related with migration of multi-planet systems in gaseous disks, and
to improve current results of migration of massive planets by
including global viscous evolution as well as detailed studies of the
co-orbital region during migration.

We perform different tests using the public code FARGO3D to validate
this method and compare its results with those obtained using a
classical fixed grid.
\end{abstract}


\section{Introduction}

Giant planets form in gaseous environments, from which they acquire
most of their mass. The interaction between both components, solids
and gas, is extremely complex and one of the key aspects in any
planetary formation theory. A planet immersed in a disk will accrete
gas and experience changes in its mass, temperature and chemical
composition. However, gravitational interactions are also important,
as they exert a force onto the planet which dramatically alters its
orbit, through an exchange of energy and angular momentum between the
disk and the planet.

The gravitational force due to a planetary embryo excites spiral waves
in the disk. These are launched at certain specific regions, known as
Lindblad resonances, which combine to form a large scale asymmetric
spiral wake co-rotating with the planet
\citep{1986Icar...67..164W,2002MNRAS.330..950O,2002ApJ...569..997R}. Additionally,
material trapped in the horse-shoe region also contributes to the
exchange of angular momentum. This so-called co-rotation torque is
typically of same order of magnitude as the torque produced by the
spiral wake \citep[and references therein]{2013LNP...861..201B}. The
net effect, considering both Lindblad and co-rotational torques,
yields a time variation of the planet's semi-major axis usually
referred to as type I migration.

The total gravitational torque between the gas disk and a low mass
planet can be estimated analytically under certain simplifications,
giving important insights about the physical mechanisms involved in
migration \citep[e.g.][]{1979ApJ...233..857G,
  2002ApJ...565.1257T,2008ApJ...672.1054B,2009MNRAS.394.2283P}. However,
in the past two decades it has been increasingly clear that numerical
experiments are indispensable in order to advance in migration
theories.

Hydrodynamical simulations are computationally demanding and extremely
costly since they must: (i) resolve the narrow region near the
planet's orbit for a proper calculation of the co-rotation torque,
(ii) maintain a good resolution far from the planet to resolve the
spiral wake, and (iii) consider the entire azimuthal extension of the
orbit to adequately describe the horseshoe motion.

Significant efforts have been made to gain in efficiency and realism,
either by developing new algorithms to improve the orbital advection
\citep{2000A&AS..141..165M} and/or by considering a self-consistent
disk evolution during migration \citep{2007A&A...461.1173C}. In
particular, this second feature has been shown to be crucial in
studies of the so-called type II migration, which occurs when the
planet is massive enough to open a gap around its horseshoe region.

Recent exoplanetary detections, particularly by the {\it Kepler}
mission, show the existence of many multi-planetary systems close to
the star, some of which are located near (but outside) mean-motion
resonances \citep{2011ApJS..197....8L,2014ApJ...790..146F}. It is yet
unclear whether these systems were formed in-situ
\citep[e.g.][]{2013ApJ...775...53H} or transported from beyond the ice
line as a result of disk-planet interactions
\citep[e.g.][]{2013ApJ...778....7B}.

Hydro-simulations of multiple planets face two challenges. On one
hand, they must be able to follow the orbital decay from initial
semi-major axes $a \sim 3$~AU down to final configurations with $a
\sim 0.1$~AU. In other words, simulations must be large-scale. On the
other hand, even after the bodies reach the region close to the
central star, their orbital evolution must be extended over a
sufficiently long time scale to allow for the resonant dynamics to act
and guarantee that the outcome actually corresponds to a stationary
solution. Thus, simulations must not only be large-scale but also
long-term.

So far, hydro-codes have not been able to comply with these
requirements. One of the main difficulties is caused by the
Courant-Friedrich-Levy (CFL) condition, a stability constrain inherent
to all explicit hydrodynamic solvers. In a cylindrical or spherical
mesh, the CFL condition generally limits the time step through the
size of the innermost cell. As a result, a gas disk that is modeled to
include the regions close to the star will have a very small step
size, even if the planets are initially located at large semi-major
axis and if an orbital advection algorithm is used. Thus, radially
extended and long-term numerical experiments have been practically
impossible, and all simulations of planetary migration have been
local, both with respect to space and time, with a few particular
exceptions \citep[e.g.][]{2007MNRAS.378.1589M,2009ApJ...705L.148C}.

In this paper we present a new numerical method that overcomes these
limitations. It basically consists of a remapping technique, which
updates the mesh boundaries of the disk according the location of the
migrating planets. The redefinition of the boundaries is automatic,
leading to high-resolution simulations that are independent of the
orbital evolution of the system. Moreover, since the mesh is not
defined in regions that are not required, the code gains in efficiency
and precision, allowing simulations virtually infinite in time and
space.

Our paper is organized as follows: in section \ref{sec:model} we
present the standard disk model used for this work, in section
\ref{sec:method} we present the remapping method, in section
\ref{sec:tests} we present different tests used to validate our
implementation. In section \ref{sec:discussion} we discuss the
benefits of our method compared with the current techniques. In
section \ref{sec:multiplanet} we present two simulations including 2
and 3 planets showing the capabilities of the method. In section
\ref{sec:conclusions}, we present a final discussion and the
conclusions of this work.

\section{Disk model}
\label{sec:model}

We describe the problem using the Navier-Stokes equations in a
non-rotating inertial frame. For simplicity we solve these equations
integrated over $z$, assuming a two dimensional non self-gravity gas
disk orbiting a central star of mass $M_*$.  Continuity equation
reads:
\begin{equation}
  \frac{\partial \Sigma}{\partial t} + \nabla.\left(\Sigma \vec{v}\right) = 0,
  \label{eq:continuity}
\end{equation}
where $\Sigma$ is the surface density, and $\vec{v}$ is the velocity
field of the fluid.  The momentum equations are:
\begin{equation}
  \label{eq:momenta}
\rho \left(\frac{\partial \vec{v}}{\partial t} +
\vec{v}.\nabla\vec{v}\right) = -\nabla P - \rho\nabla\vec{\phi} +
\nabla . \vec{\Pi},
\end{equation}
where $P$ is the pressure and $\phi$ is the gravitational potential,
which includes the contribution of the star and planets.  $\vec{\Pi}$
is the stress tensor, given by:
\begin{equation}
  \vec{\Pi} = \rho \nu \left[ \nabla \vec{v} + \left(\nabla \vec{v}
    \right)^T - \frac{2}{3}\left(\nabla.\vec{v}\right) \vec{I}
    \right],
\label{eq:stresstensor}
\end{equation}
with $\nu$ the kinematic viscosity and $\vec{I}$ the identity tensor.

We use a locally isothermal equation of state, in which the gas
pressure is related to the surface density by $P = c_s^2(r)\Sigma$,
with $c_s$ the sound speed of the gas.  $c_s$ is related with the
aspect ratio $h$ of the disk by $c_s = r\Omega_k h(r)$, which is given
in term of the vertical disk scale height $H$ by $h(r) = H(r)/r$. The
Keplerian angular velocity is $\Omega_k$, given by $\Omega_k =
\sqrt{GM_*/r^3}$, where $G$ is the gravitational constant and $r$
denotes the distance to the central star.

In our model, we adopt power laws for $\Sigma$ and the aspect ratio
$h$ of the disk, with indices $\alpha$ and $f$ respectively:
\begin{equation}
\label{eq:1}
  \Sigma \propto r^{\alpha} \mbox{~~~and~~~} h\propto r^f.
\end{equation}
This implies a power law for the temperature $T$, with index
$\beta=2f-1$.  The planet of mass $m_p$ orbits the central star and
feels the gravitational potential of the disk.  It has a softened
potential of the form:
\begin{equation}
\phi_p = -\frac{GM_*}{\sqrt{\left|r-r_p\right|^2+\epsilon^2}}
\label{eq:potential}, 
\end{equation}
where $\epsilon$ is the softening length. We use a softening length
parametrized by the pressure scale height in the form $\epsilon =
0.6H$.  We neglect the indirect terms since they are not relevant to
our concerns.

Damping boundary conditions are applied following
\citet{2006MNRAS.370..529D}. These are required to match the
large-scale and long-term behavior of the disk, given either by power
laws or by an auxiliary one-dimensional calculation. We use the
hydrodynamical code FARGO3D \citep{0067-0049-223-1-11} over a polar
mesh evenly spaced in radius and azimuth.

\section{Method and Numerical implementation}
\label{sec:method}

The main idea behind our method is to take a previous state as a
spatially re-sampled initial condition to advance the system into a
new state.  There are two possible ways to do this.  One way is to
integrate the system to a specific state and then re-sample the
physical quantities and restart the simulation.  A second possibility
is to use a continuous recipe, where for each time step the mesh is
re-sampled.  We choose the second option since it allows for smooth
changes of all quantities.

Since we are interested in following planets while they migrate, the
re-sampling must be performed exclusively radially.  The algorithm we
developed has three basic steps, consisting of a suitable recipe to
calculate the new radial borders of the mesh, a method to re-fill the
mesh with appropriate values for the hydrodynamic quantities and,
finally, an update of the boundary conditions and damping zones.

To find the radial borders for the new mesh we use the planets'
location.  The gas ring furthest from the planet that is able to exert
a torque is located at the inner (outer) 2:1 (1:2) mean motion
resonance (MMR) with the planet \citep{1979ApJ...233..857G}, which are
the furthest Lindblad resonances with respect to the planet.  Thus,
the new border positions should be calculated as a function of MMRs or
period ratios with the planet closest to them.  In this way we ensure
to incorporate all the ingredients of type I migration during the
entire simulation.

Once we have computed the new border positions, we radially split the
mesh using the same number of cells than in the previous step.  This
choice is motivated by the simplicity of implementation.

Finally, we perform a linear interpolation, using the old fields, to
calculate the values at the position of the new cells.  If the new
cells fall outside the old domain, we extrapolate the fields using
analytical prescriptions given either by initial conditions or by an
auxiliary viscous evolution model (see \S \ref{subsec:viscous}).

\begin{figure}
\centering
\includegraphics[width=\columnwidth]{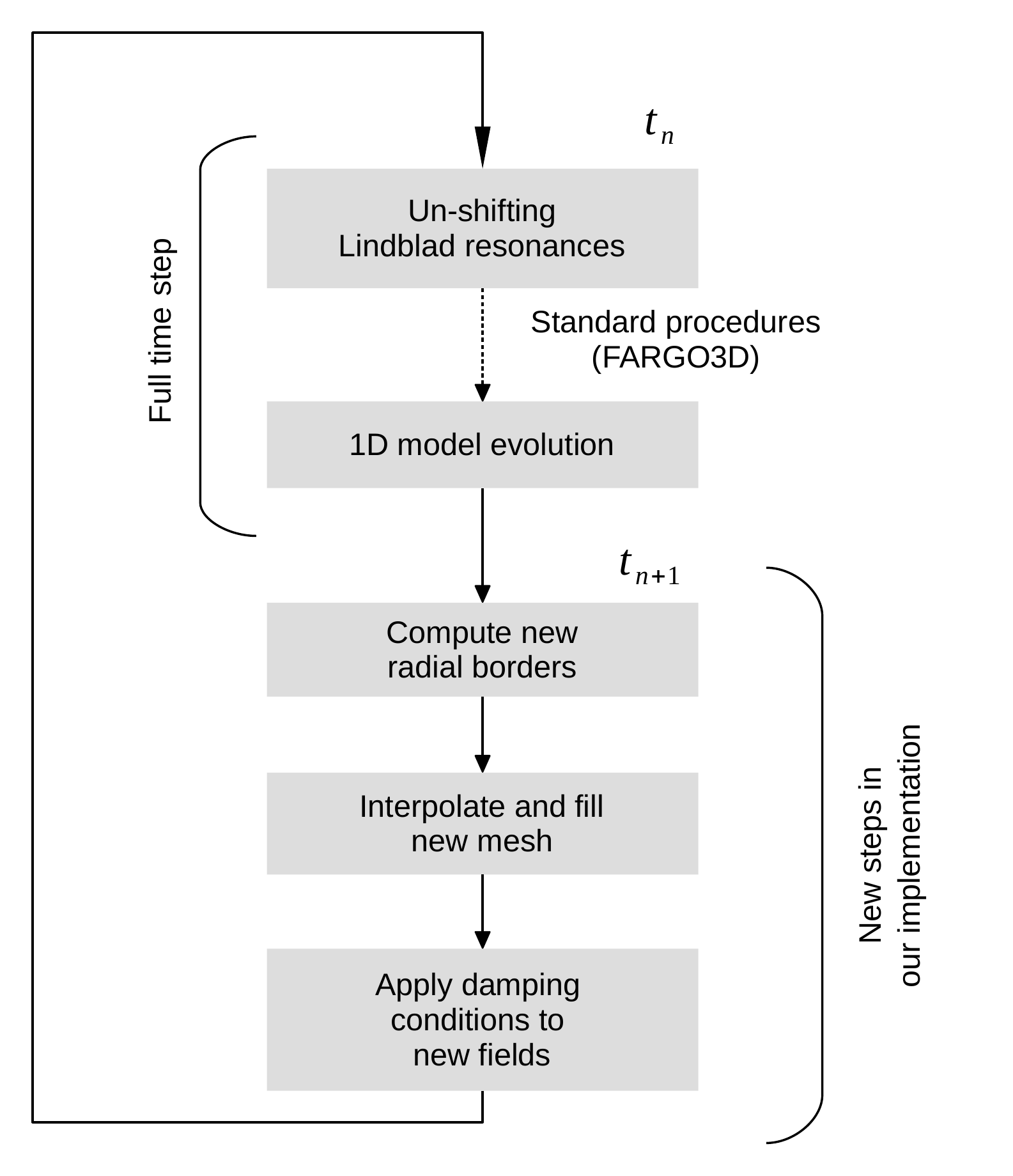}
\caption{\label{fig:fig_algorithm} Flow chart showing how our
  implementation is coupled to the standard flow of FARGO3D. For
  simplicity we show only the part corresponding to a full update (see
  figure 2 in \cite{0067-0049-223-1-11} for details). Inside the main
  loop, we follow the method described in \S \ref{subsec:unshift} for
  un-shifting the Lindblad Resonances and we use the time step already
  calculated by the CFL condition to advance the 1D model for the
  surface density under the effect of the viscous evolution of the
  disk (see \S \ref{subsec:viscous}). After that, a complete time step
  is performed for all fields. Then, we recompute the new location of
  the borders of the mesh. Finally, before advancing another time
  step, we update the damping zones. All fields are damped toward the
  updated reference values, given by the prescriptions detailed in \S
  \ref{subsec:filling}.}
\end{figure}

In Fig.~\ref{fig:fig_algorithm} we present a simple flow chart showing
how the remapping algorithm is implemented.

\subsection{Defining the new borders}
\label{subsec:borders}

\begin{figure}
\centering
\includegraphics[width=\columnwidth]{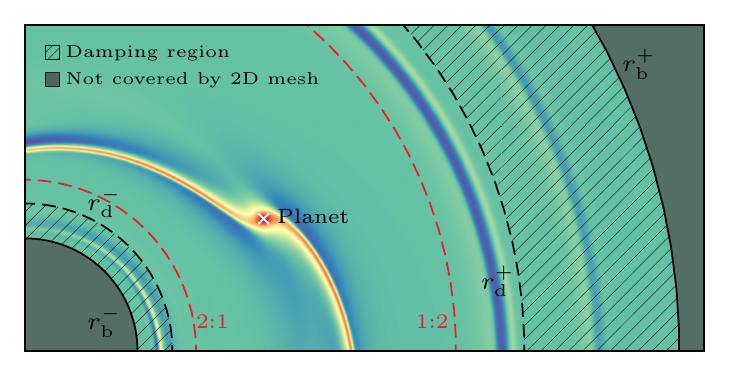}
\caption{\label{fig:fig_mesh} A wedge of a typical perturbation of
  surface density. The gray zone marks the region of the disk where
  the 2D mesh is undefined. Black continuous lines represent the
  border of the active mesh.  The hatched zones show where damping
  boundary conditions are applied to the fields. They are bounded by
  black dashed lines. The mesh portion that extends from the inner to
  the outer black dashed lines is referred as the active mesh.  In the
  case shown here, we set up the dashed lines using
  Eq.\eqref{eq:damping} with $\mathcal{R}_\text{d} = 5/2$. For damping
  zones, the distance from the first ring to the corresponding border
  of the mesh is computed using Eq.\eqref{eq:border1} with
  $\mathcal{R}_\text{b} = 3/2$.  In red dashed lines we show the
  location of the 2:1/1:2 MMRs with the planet, which are important
  locations because only between them the gas exert an effective
  torque onto the planet, marked by a white cross.}
\end{figure}

To define the borders for the mesh we have to consider two different
zones: (i) the active mesh (where there is no damping), and (ii) the
damping zones.

The active mesh is defined in terms of period ratios with respect to
the planets located in the mesh:
\begin{equation}
r_{\text{d}}^{\pm} = r_p^{\pm} \mathcal{R}_{\text{d}}^{\pm 2/3},
\label{eq:damping}
\end{equation}
where $r_p^{\pm}$ is the semi-major axis of the closest planet to the
$\pm$ border and $\mathcal{R}_{\text{d}}$ is a positive real number
greater than one, which represents the period ratio between the planet
and the gas at the border of the active mesh (without considering the
damping zones).

Once we have determined the location of the active mesh, we proceed to
calculate the total extension of the mesh. Since the characteristic
time for damping is normally chosen proportional to the local orbital
period of the damping cells \citep{2006MNRAS.370..529D}, it is natural
to define its width so as to span a fixed extent in orbital periods,
in the same way as was done to define $\mathcal{R}_{\text{d}}$:
\begin{equation}
r_{\text{b}}^{\pm} = r_{\text{d}}^{\pm} \mathcal{R}_{\text{b}}^{\pm 2/3},
\label{eq:border}
\end{equation}
where $\mathcal{R}_{\text{b}}$ is a positive real number greater than
one, which represents the period ratio between the first damping ring
and the border of the mesh.

Using Eqs.~\eqref{eq:damping} and \eqref{eq:border} we obtain the
location of the borders that keeps a constant size in periods with
respect to the planet, given by:
\begin{equation}
r_{\text{b}}^{\pm} = r_p^{\pm} (\mathcal{R}_{\text{d}}\mathcal{R}_{\text{b}})^{\pm 2/3}.
\label{eq:border1}
\end{equation}

In Fig.~\ref{fig:fig_mesh}, we show a sketch of a wedge for a typical
simulation.  The damping regions are represented by the hatched zones
and their limits indicated by dashed black lines, whose location is
determined by Eq. \eqref{eq:damping}.  The limits of the 2D mesh
(black continuous lines) are given by Eq.~\eqref{eq:border1}.  Red
dashed lines show the location of the 2:1/1:2 MMR between gas and the
planet.  In this particular case, we have chosen the values
$\mathcal{R}_{\text{b}}=3/2$ and $\mathcal{R}_{\text{d}} = 5/2$.

\begin{figure}
\centering
\includegraphics[width=\columnwidth]{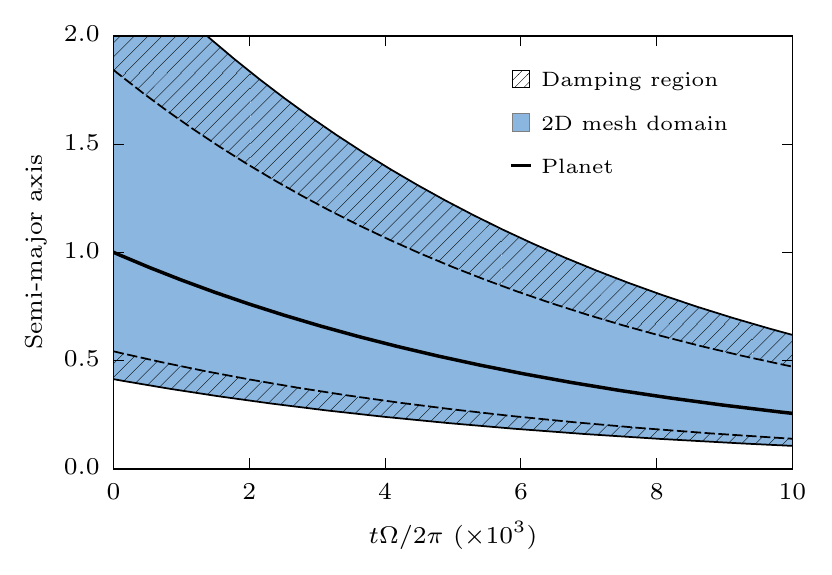}
\caption{\label{fig:fig_a} Semi-major axis evolution of a planet
  (thick black line) using the remapping method. The blue region
  correspond to the extension of the 2D mesh. Damping regions are
  hatched.}
\end{figure}

Fig.~\ref{fig:fig_a} shows the evolution of the semi-major axis of a
low mass planet for a typical simulation. The mesh is modified
continuously at runtime to keep the borders at a fixed period ratio
with the planet, so as to track it. The mesh is shown by the blue
region, and its extension can be calculated by using
Eq.~\eqref{eq:border1}.

\subsection{Filling the mesh}
\label{subsec:filling}
Once we have computed the new borders, we radially split the mesh
using the same number $n_r$ of radial cells as in the previous state.
In this work we use an even spacing between cells, but any other
spacing is possible, such as a logarithmic spacing.  A further
discussion on this choice is given in \S \ref{subsec:spacing}.

For each new cell, we calculate its neighbors on the old mesh and use
them to estimate the new values using linear interpolation.

If two neighbors are not available, or if the updated cell falls onto
the damping rings, we use analytical prescriptions to compute the
corresponding values.  These prescriptions may be given by the initial
conditions for the sound speed (or temperature) and the azimuthal
velocity.  Similarly, the initial conditions are used to prescribe the
surface density and radial velocity if the viscous evolution of the
disk is not taken into account; otherwise, as we will see in \S
\ref{subsec:viscous}, we use Eq.~\eqref{eq:pringle} to update the
surface density and Eq.~\eqref{eq:vr} to get the radial velocity.

Now we will discuss the basic algorithm for the case in where the mesh
is linearly spaced.

We start from a quantity $q$ at location $r_i$ at time $t_n$, and
denoted by $q_i^{n}$. After a time step, the new updated value
$q_i^{n+1}$ is obtained by solving the hydrodynamics equations using
finite difference upwind, dimensionally split methods, combined with
the FARGO algorithm \citep{2000A&AS..141..165M} for orbital advection
and a fifth order Runge-Kutta integrator for advancing the planets
\citep{0067-0049-223-1-11}.

\begin{figure}
\centering
{\includegraphics[width=\columnwidth]{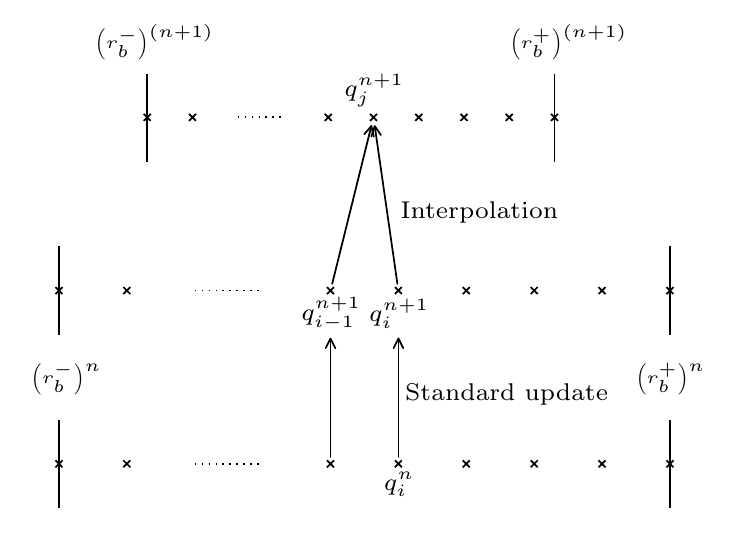}}
\caption{\label{fig:algorithm} Diagram showing how a quantity
  $q_j^{n+1}$ in the remapped mesh is obtained from old values.  In
  the bottom part we start at time $t_n$ and update the quantity
  $q_i^n$ to a new value $q_i^{n+1}$ (middle step) by evolving the
  hydrodynamics equations. Finally, at the top part we compute the new
  borders of the mesh and we fill it at location $r_j$ by
  interpolating the recently updated field.}
\end{figure}

The border locations are updated, from
$\left(r^-_b\right)^{n},\left(r^+_b\right)^{n}$ to
$\left(r^-_b\right)^{n+1},\left(r^+_b\right)^{n+1}$ following the
recipe described by Eq.~\eqref{eq:border1}. After that, we need to
compute the new radial locations $r_j$ for each cell. In the evenly
(or linearly) spaced case, this can be done using the simple relation
$r_j = (r^-_b)^{n+1} + j\Delta $, $j=0...n_r-1$, and $\Delta =
\left[(r^+_b)^{n+1}-(r^-_b)^{n+1}\right]/n_r$. For the sake of
definiteness, we will assume that $q$ is a face-centered (\emph{i.e.}
staggered) quantity, such as $v_r$, the radial velocity.  Its value
after remapping is given by:
\begin{equation}
q_j^{n+1} = q_{k_j}^{n+1} +
\left.\frac{q_{k_j+1}-q_{k_j}}{r_{k_j+1}-r_{k_j}}\right|^{n+1}(r_j-r_k),
\label{eq:cant}
\end{equation}
with $k_j$ a function of the old spacing. In the linearly spaced case,
the index $k_j$ is given by:
\begin{equation}
k_j = \text{int}\left[\frac{r_j - (r_b^-)^n}{(r_b^+)^{n} - (r_b^-)^n }
  ~ n_r \right],
\end{equation}
where $\mathrm{int}(x)$ represents the largest integer inferior to
$x$.  The extension to the case of cell-centered variables is
straightforward. In Fig.~\ref{fig:algorithm} we depict this algorithm,
and we put in context all important quantities.

\subsection{Viscous evolution of the disk}
\label{subsec:viscous}

\begin{figure}
\centering
{\includegraphics[width=\columnwidth]{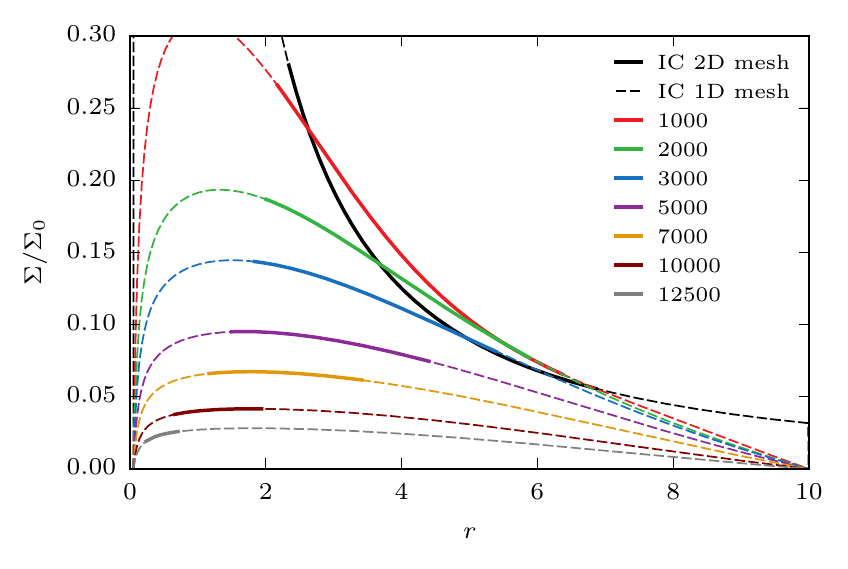}}
\caption{\label{fig:visc_evolv} Viscous evolution of a surface density
  profile initially given by a power law $\Sigma\propto r^{-1.5}$ with
  no planet.  The number of cells used for the 1D viscous model
  (Eq.~\ref{eq:pringle}) is 2048 while the 2D mesh has 256 radial
  zones.  The viscosity $\nu$ is constant over the entire domain and a
  zero surface density boundary condition is applied at $r=0.05$ and
  $r=10$.  The 2D mesh has an imposed, constant migration rate.
  Labels are given in orbital periods at $r=1$.}
\end{figure}

Since the remapping method allows to perform long-term simulations
over a wide spatial range, special considerations regarding the global
viscous evolution of the disk should be taken into account.
Describing a disk with a power law for long-term simulations may not
be an appropriate choice.  From the third term in the right hand side
of Eq.~\eqref{eq:momenta}, the entire disk is expected to drift over a
time scale $\tau_\nu \propto r^2/\nu$ \citep{1981ARA&A..19..137P}.  On
the other hand, the characteristic time scale for migration ($\tau_a =
r_p/\dot{r_p}$) in the type I regime scales as $\tau_a \propto
M_*^2(h/r)^2 (m_p\Sigma)^{-1} \Omega^{-1}$ \cite{1986Icar...67..164W}.

Thus, the ratio between both time scales is given by:
\begin{equation}
\frac{\tau_a}{\tau_\nu} \propto \frac{M_*^2h^2\nu}{m_p \Omega \Sigma r^4}.
\label{eq:timesratio}
\end{equation}
At $r=R_0=5.2$~AU, typical parameters are $h = 0.05$, $\nu=10^{-5}
R_0^2\Omega_0$ and $\Sigma \sim 6\cdot 10^{-4} M_\odot R_0^{-2}$, and
assuming $M_* = 1~M_\odot$, $m_p = 10^{-5}~M_\odot$ it gives a ratio
equal to $\sim$ 4.  This value depends on $r$ as $r^{-(2.5+\alpha)}$
(assuming $f=0$), where $\alpha$ is defined by Eq.~\eqref{eq:1}, and
in practice is a negative number greater than $-1.5$.  Thus, the
migration rate becomes negligible compared to the viscous time as the
planet migrates to the star.  This simple analysis shows that in order
to perform a consistent simulation of migrating planets in type I
regime over a long scale (in space and time), a self-consistent
viscous evolution of the disk should be considered.  We address this
issue by solving a simplified one-dimensional model for the angular
momentum conservation of the disk under the effects of viscosity and
neglecting gas pressure and planet torques
\citep{1981ARA&A..19..137P}:
\begin{equation}
\frac{\partial \Sigma}{\partial t} = \frac{3}{r}
\frac{\partial}{\partial r} \left[\sqrt{r}\frac{\partial}{\partial
    r}\left(\nu \Sigma \sqrt{r}\right)\right].
\label{eq:pringle}
\end{equation}
We evolve Eq.~\eqref{eq:pringle} explicitly over a sufficiently large
one-dimensional radially extended mesh at runtime during the full
two-dimensional simulation.  We use the advanced surface density to
update the boundaries at the damping regions.  Once the surface
density is calculated, the radial velocity can be inferred by the
closed relation:
\begin{equation}
v_r = -\frac{3}{\Sigma \sqrt{r}}\frac{\partial }{\partial r}\left(\nu
\Sigma \sqrt{r}\right).
\label{eq:vr}
\end{equation}

This implementation of the viscous evolution for the disk by using a
one-dimensional mesh can be compared to the implementation of
\cite{2007A&A...461.1173C}.  In our case we treat reflections by using
the well known damping zones \citep{2006MNRAS.370..529D} and we do not
address the angular momentum conservation issue because we assume that
planets do not strongly alter the surface density far away from its
current location. This approach is not as sophisticated as that of
\cite{2007A&A...461.1173C}, but we will see in next sections that we
can achieve very clean and accurate results.  One important difference
with their implementation is that the 2D mesh was fixed, so that
planets could get close to the boundaries and some care had to be
taken with the wake's momentum flux into the one-dimensional
grid. Here our prescription of Eq.~\eqref{eq:border1} allows planets
to stay clear of the boundaries, and our damping boundary prescription
is sufficient to ensure a smooth transition between the 2D and 1D
meshes.

Fig.~\ref{fig:visc_evolv} shows an example of how the 2D migrating
mesh (using the remapping method) can be altered by considering
long-term simulations.  In this figure can be compared each of the
continuous lines with the expected solution given by the corresponding
dashed line. This figure corresponds to the test shown in \S
\ref{subsect:viscous_evolv}.

\subsection{Un-shifting Lindblad resonances}
\label{subsec:unshift}

\begin{figure}
\centering
{\includegraphics[width=\columnwidth]{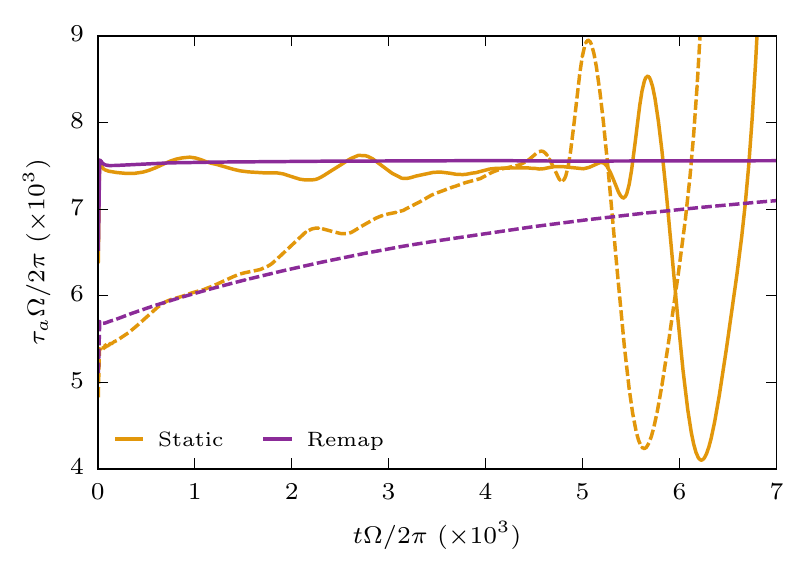}}
\caption{\label{fig4} Characteristic migration time for the static
  (orange) and remapping (purple) cases. Continuous and dashed lines
  correspond to results with and without correcting for the spurious
  shift of Lindblad resonances respectively. In agreement with
  \cite{2008ApJ...678..483B}, the migration time scale is shorter when
  the planet and the disk do not orbit the same potential.}
\end{figure}

\cite{2008ApJ...678..483B} show that a special treatment for computing
the torque is needed when self-gravity is not included in simulations
of migrating planets.  They show that, due to a spurious shift of
Lindblad resonances, a migrating planet in a non-self gravity disk
migrates faster than theoretical predictions and in fully self-gravity
simulations.

This spurious shift occurs because planet and disk orbit a different
effective potential, since the planet feels the disk gravity but the
disk does not feel itself.

The spurious resonance shifting can be easily addressed by subtracting
the azimuthal mean density prior to the force calculation.  This
method does not impact on the net torque, but changes the orbital
frequency of the planet, fixing the shift and ensuring an accurate
migration rate.

To compute the gravitational force first, we subtract the mean surface
density at time $t_n$:
\begin{equation}
	\delta \Sigma(r,\varphi) = \Sigma(r,\varphi) -
        <\Sigma>_\varphi,
	\label{eq:meanback}
\end{equation}
where $\varphi$ is the azimuthal coordinate and $<\Sigma>_\varphi$ is
defined as:
\begin{equation}
 <\Sigma>_\varphi = \frac{1}{n_\varphi}\sum_{i=0}^{n_\varphi - 1}
 \Sigma(r,\varphi_i),
\end{equation}
and we calculate the gravitational force using $\delta \Sigma$.

To our concerns, this shift is extremely important to obtain
convergent migration rates, as can be seen in Fig.~\ref{fig4},
corresponding to the test of \S~\ref{subsec:fixingtest}.  By
considering different radial extensions for the mesh (at a fixed
resolution), we are artificially including a different amount of mass
in the inner regions of the disk. This changes the planet's orbital
frequency with respect to the disk and directly impacts the net torque
exerted onto the planet. This effect is clearly shown by the purple
dashed line of Fig.~\ref{fig4}, where the asymptotic value,
corresponding the a narrow and low mass mesh, tends to its respective
fixed case (purple continuous line).

\section{Numerical tests}
\label{sec:tests}

In this section we show a set of tests to validate the remapping
method.  Runs presented here include all the features already
described: (i) the remapping method, (ii) the global viscous evolution
and (iii) the shift to the Lindblad resonances.  For comparison
purposes we do not include a certain feature in some tests, like the
shifting for the Lindblad resonances and the remapping method for
fixed-mesh simulations.

We used the public code
FARGO3D\footnote{\href{http://fargo.in2p3.fr/}{http://fargo.in2p3.fr/}}
adapted to include our method using the disk model presented in \S
\ref{sec:model}.  Our mass unit is $M_* = 1$, our length unit is
$R_0$, the planet's initial orbital radius, and our time unit is $t =
\Omega_k^{-1}$ (in this unit system we have therefore $G=1$). We
consider a planet of mass $m_p = 2\times 10^{-5}M_*$ and a gaseous
disk of parameters $\Sigma(R_0)=10^{-3}$, $\alpha=-0.5$, $h(R_0) =
0.05$, $f=0$. We use the prescription of \cite{1973A&A....24..337S}
for viscosity, defined by $\nu = \alpha_\nu H^2\Omega$, with a
constant value $\alpha_\nu=4\times 10^{-3}$.  The azimuthal domain is
$[0,2\pi]$. The radial domain will be described for each particular
case.  In the tests without remapping, the damping regions span a
period extension equal to $3/2$ from each active border, using a
prescription similar to that employed for the remapping case
(Eq. \eqref{eq:border}).

\subsection{Results without the remapping technique}
\label{subsec:teststatic}

In this section we measure the migration rate of a single planet,
considering a static (not remapped) disk. Both the borders and damping
regions are assumed fixed. An important parameter involved in this
test is the characteristic damping time. By numerical experiments we
found that a good value to minimize reflections considering the
adopted extension for the damping zone is 0.3$\Omega^{-1}$. The limits
of the mesh were chosen as $R_{min} = 0.37$ and $R_{max}=2.08$, and we
consider three different resolutions: $n_r\times n_{\Phi} = 462\times
512$, $924\times 1024$ and $1386\times 1536$.

\begin{figure}
\centering
\includegraphics[width=\columnwidth]{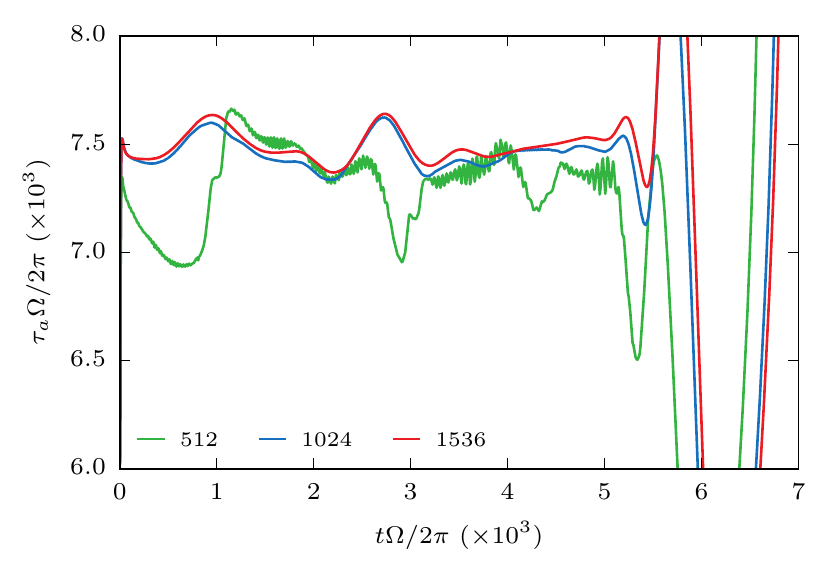}
\caption{\label{fig1} Characteristic migration time $\tau_a$ for three
  different simulations considering a static mesh (without the
  remapping method) but considering different resolutions: the green
  continuous line correspond to the lower resolution case, the blue
  continuous line doubles it and the red one increases it by a factor
  3.  The three curves shown a similar overall behavior. Higher
  oscillations are present in the lower resolution case. At time
  $\sim5.5 \times 10^{3}$ orbits curves diverge because the planet
  reach the inner limit of the mesh, and migration halts. As is
  described in the main text, oscillations can be attributable to
  reflections of the wake on both boundaries.}
\end{figure}

In Fig.~\ref{fig1} the evolution of the characteristic migration time
for the three different cases is shown. A converged result is fairly
achieved by the middle resolution case, as is shown by the blue and
red curves.

We observe very small oscillations in the migration rate for the three
cases, which don't depend strongly on resolution and can be explained
as a reflection of the wake by the borders of the mesh.  We have
checked this assumption by performing simulations with different
radial extension but at the same resolution. We have observed a shift
of these oscillations when the radial extent changes.  As we will see
in the next sections, these oscillations are not present when we use
the remapping method for an isolated planet and a non-flared disk
(e.g. purple lines of Fig.~\ref{fig4}). This does not mean that the
remapping method allows to improve the value of the migration rate
with respect to the standard case for a comparable resolution. It only
means that the reflected wake maintains a fixed position with respect
to the planet for the entire simulation. This assertion can be easily
proved using the expression for the azimuthal position of the wake
provided by \cite{2002MNRAS.330..950O}. According to these authors,
the azimuthal angle of the wake scales linearly with the distance of
the planet and $h$. Since our method sets the edges at a distance
proportional to the planetary orbital radius, this angle remains
constant.

Finally, from the amplitude of these oscillations, we can get an idea
of the precision for the measured migration rates (not better than
$3\%$ in this case, which is extremely good). In principle, the
precision achieved depends on how well reflection can be avoided, so
we could modify the damping prescription to further decrease
reflection and thus improve the measures.

At time $\sim5.5\times10^{3}$ orbits all curves reach the inner
boundary of the mesh, migration halts and migration rates diverge.

\subsection{Convergence tests with the remapping algorithm}

\begin{figure*}
\centering
\includegraphics[width=\textwidth]{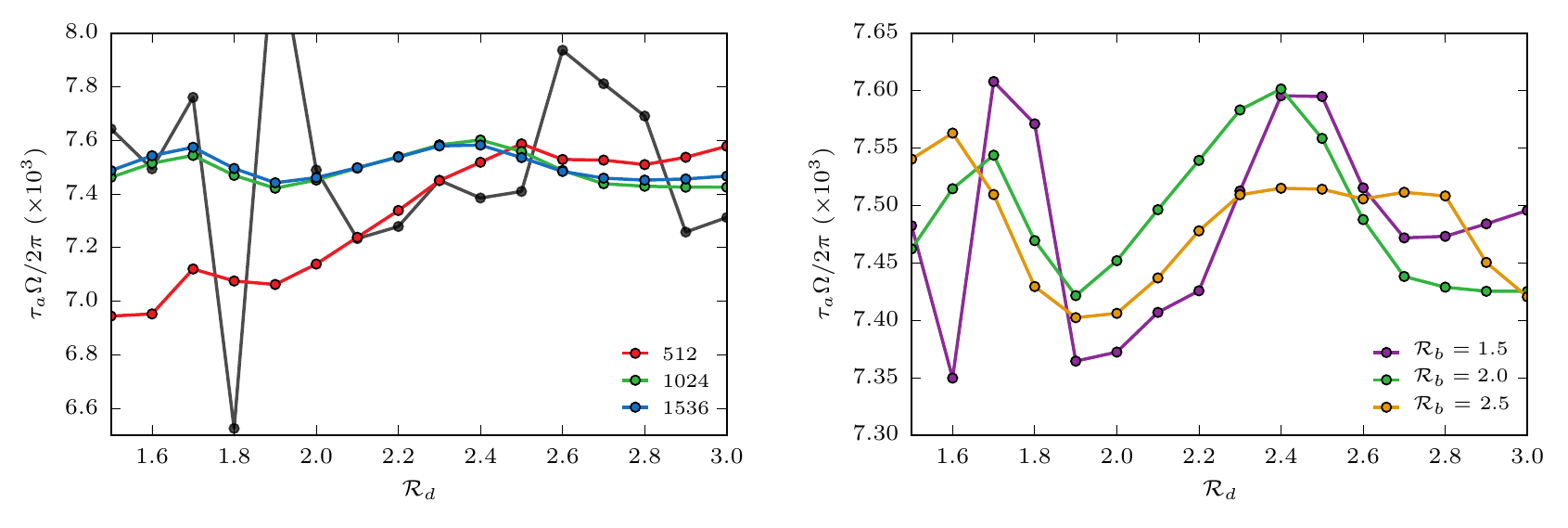}
\caption{\label{fig:f8} Characteristic migration time as a function of
  $\mathcal{R}_{\text{d}}$ for the remapping case. The left panel
  shows a set of simulations with three different resolutions
  (referred to as 512, 1024 and 1536) and same value
  $\mathcal{R}_{\text{d}}=2$. The black curve shows a set of
  intermediate resolution with a stronger damping. The right panel
  shows a set of simulations with three different values for
  $\mathcal{R}_{\text{b}}$ (1.5, 2.0 and 2.5) for the intermediate
  resolution set. It is noticeable how the amplitude of the curves
  decreases for larger $\mathcal{R}_{\text{b}}$.  }
\end{figure*}

The remapping algorithm needs two different radial extents to be
defined: (i) the active mesh and (ii) the damping zones, controlled by
the dimensionless parameters $\mathcal{R}_\text{d,b}$. In this section
we perform two tests to verify how well migration rates are measured
by varying both of them. In all cases the characteristic time for
damping was chosen as $0.3\Omega_k^{-1}$.

The first test consists in simulating an isolated planet in a disk
with the same parameters already described, but varying both the
effective resolution and $\mathcal{R}_\text{d}$ on a grid. We used
three different resolutions, referred to by the number of azimuthal
zones: $512, 1024, 1536$, corresponding to the same resolutions used
in the simulations performed without the remapping algorithm. We
varied $\mathcal{R}_\text{d}$ linearly from $1.5$ to $3$ by steps of
$0.1$. We have run each test for at least $10^3$ orbits, and the
values reported here are an average of the characteristic migration
time over the last hundred of orbits (which is constant for these
parameters).

The left panel of Fig.~\ref{fig:f8} shows the results of this
test. The lower resolution case (in red) shows a systematic increase
of the migration time, evidencing that this resolution is not
sufficient to resolve the physical processes involved. The
intermediate and high resolution cases (green and blue respectively)
display a nearly converged rate.  In this test we can see that the
oscillations observed in the static case are still present when we
measure the migration rate by considering different radial extents of
the mesh. This feature is an evidence of reflections of the wake on
the boundaries, producing a variable migration rate. The amplitude of
these oscillations gives an estimate of the precision achieved (not
better than $3\%$ in this particular case). For all runs presented in
this panel we have used $R_b = 1.5$. Surprisingly, we do not observe
an important variation of migration rate by considering
$\mathcal{R}_\text{d} < 2.0$, probably due the smooth and large
damping regions used. This parameter should nevertheless be greater
than $2$ to capture all Lindblad resonances. To test this idea we have
run a set of similar simulations but defining a very strong damping
condition over a narrow region, showed by the black curve. In this
case we can see fast and strong variations in the migration rate for
$\mathcal{R}_\text{d} < 2.0$. For $\mathcal{R}_\text{d} > 2.0$ we
still observe oscillations with lower amplitudes, attributable to the
reflection of the wake as we show below.

The second test was designed to prove that oscillations observed in
the migration rate for different values of $\mathcal{R}_\text{d}$ are
consistent with reflections of the wake. To prove this statement we
have varied the value of the parameter $\mathcal{R}_\text{b}$, taking
1.5, 2.0 and 2.5 for the intermediate resolution run (green curve in
left panel of Fig.~\ref{fig:f8}). The results are presented in the
right panel of Fig.~\ref{fig:f8}.
It is noticeable how all curves return similar values for the mean
migration rate, but the extreme values are shifted. Another
interesting feature is that the amplitude of oscillations decreases
when $\mathcal{R}_\text{b}$ increases. This is natural if we assume
that these oscillations are due to reflections, since larger damping
regions naturally better damp perturbations.

It is important to note that these tests have been done for a planet
on a circular orbit. In the case one considers planets with a finite
eccentricity, the size of the active mesh should be validated by a
similar convergence test, since in such cases higher order Lindblad
resonances (such as the 3:1 MMR), located even further from the
planet's orbit, may play a non-negligible role in the planetary
migration. Also, it could be important in some specific cases to take
into account the physical size for the resonances, which have an
effective width that should not be neglected. This size depends on the
azimuthal wave number and reaches a maximum for low values
\citep{1993ApJ...419..155A}.

\subsection{Viscous evolution test}
\label{subsect:viscous_evolv}

We carried out a viscous evolution test for an axi-symmetric disk with
no planet.  We initialize a non-flared ($f=0$) locally isothermal and
power law surface density disk. We have used a uniform viscosity $\nu
= 10^{-4}~R_0^2\Omega_0$.

The 1D viscous model, given by Eq.~\eqref{eq:pringle}, is evolved
using 2048 uniform radial zones between $r=0.05$ and $r=10$, with null
surface density boundary conditions.  The 2D mesh has 256 radial zones
and the full radial extent spans over 3 periods in space
($\mathcal{R}_\text{d,b}=1.5$). We radially move the mesh at a
constant speed to test how the viscous method couples with the
remapping one.

Fig.~\ref{fig:visc_evolv} shows the surface density evolution at
different times.  In continuous lines 2D surface density is
shown. Dashed lines correspond to the solution of
Eq.~\eqref{eq:pringle}, used to update the boundaries and damping
zones, for different times.  In black, initial conditions are shown,
while the evolution of the surface density, under the effects of
viscosity, is presented in colored lines.

We have also checked that we are able to obtain a steady solution by
considering an $\alpha_\nu$ disk model, with $\alpha=-0.5$ and $f=0$,
configuration used in all the tests considering planets.

Results obtained are remarkably clean, allowing to perform large-scale
migration of planets considering the global viscous disk evolution,
and even open the possibility for long-term simulations of type II
migration.
 
\subsection{Fixing orbital frequencies}
\label{subsec:fixingtest}

An important characteristic that we have included in our method is the
correction for the Lindblad resonances spurious shifting.  It is a
crucial component for having converged simulations by using the
remapping method, as was already described in \S \ref{subsec:unshift}.

We have checked our implementation by using two different simulations
set, one without the remapping method and the other one by considering
all the ingredients. The parameters used are the same as those used in
the fixed cases (see \S \ref{subsec:teststatic}), but in the remapping
case we consider $\mathcal{R}_\text{d} = 2.0$ and
$\mathcal{R}_\text{b}=1.5$. Both cases have same initial resolution.

We have computed the gravitational force including the mean background
density and subtracting it, following Eq.~\eqref{eq:meanback}.

Fig.~\ref{fig4} shows the evolution for the characteristic migration
time for these four simulations. In orange lines we show the
simulations that do not include the remapping method while purple
lines show the simulations that include it. Dashed lines correspond to
the non-corrected cases and continuous lines correspond to the fixed
ones.

As we evidenced in section \ref{subsec:unshift}, and in agreement with
\cite{2008ApJ...678..483B}, the shifted cases (non-corrected) have a
shorter migration time scale.

It is interesting to note that the non corrected case using the
remapping method, has a non zero slope. It is possible to prove by
linear calculations \citep[for instance]{2002ApJ...565.1257T}, that
this slope should be in fact flat for this parameter set. Thus, it
clearly shows that migration rate achieved without the correction is
not right. The same occur for the static case.

Another important feature of Fig.~\ref{fig4} is that both, the static
and the remapping cases tend toward the corrected one at larger
time. This is a consequence of having a narrow and virtually massless
disk in the inner regions.

Finally we remark that in the remapping case we do not observe
oscillations for the migration rate (for a one planet case). This
feature can be explained as a constant shift between the reflected
wake and the planet, as was already discussed in
\S~\ref{subsec:teststatic}.

\subsection{Resolving the disk pressure scale}
\label{subsec:spacing}

A difficulty with the remapping method in the form presented here is
the requirement of maintaining an overall sufficient resolution for a
given disk model.

The pressure scale length of the disk should be well resolved at least
near the planet, which also implies to resolve the closest Lindblad
resonances, accumulating at $r\simeq \pm H$
\citep{1993ApJ...419..155A}.

In tests presented in this paper we have only considered a non-flared
disk, implying $H\propto r$.  In an evenly spaced mesh, a uniform
sampling remapping method has a linearly increasing resolution with
$r_p$, and the number of cells per unit of $H$ will be the same for
the entire simulation, thus ensuring converged results for migration
rates.  However, when we consider a flared disk, the number of cells
per unit of $H$ is $\propto r^{-f}$, which is a decreasing function
for flared disks ($f>0$). Thus, a growing number of cells per unit of
$H$ is needed when the planet migrates to the star.

This highlights the need for non-uniform grids to simulate migrating
planets in a flared disk, which ideally should be radially spaced as
$\propto r^{f+1}$.

However, in real cases disks can not be flared over large radial
range, limiting the maximum flare allowed. Thus, in practice a
sufficiently resolved logarithmic migrating mesh coupled with a
realistic disk model, like a non-global power law for density and or
temperature, should be good enough to capture all important details of
planetary migration for large-scale migration.

\section{Improvements and conservation properties}
\label{sec:discussion}

If quantities are power laws of radius, such as the surface density,
the temperature and the rotational velocity, a procedure to minimize
errors during interpolations could be to perform them in a log-log
space (which is actually an exact reconstruction in the unperturbed
case).  It can be done by defining $\chi=\log(r)$, $\xi = \log(q)$ and
replacing $r\rightarrow \chi$ and $q\rightarrow\xi$ in
Eq.~\eqref{eq:cant}.  However, all results presented in this work were
performed using interpolations in the linear space. For comparison
purposes we have run a few of them using a logarithmic interpolation
method for quantities defined as a power law, showing minute
differences ($\sim0.2\%$) in the measured migration rates over the
whole time.  In more realistic cases, in which the disk cannot be
modeled as a global power law, we don't expect to have a significant
difference between linear or logarithmic interpolation.

Another method for un-shifting the Lindblad resonances was proposed by
\cite{2008ApJ...678..483B}, and consists in adding to the star
potential the potential of an axi-symmetric disk that has same density
profile as the disk. This method is slightly more complex to
implement, and in theory slightly more accurate than the one used in
this work, because it matches more closely the real orbital
frequencies. Nonetheless the method used here is largely sufficient to
obtain migration rates within a few percent accuracy.

Regarding conservation properties, our method cannot ensure
conservation of mass (and consequently neither angular momentum) since
a rescale of the cells plus an interpolation procedure alters the
amount of mass inside a given cell. Despite of that, the total error
accumulated during a complete simulation by performing consecutive
interpolations is actually small. Our tests show that the total error
(compared with analytic expectations) in the worst cases, such as one
with low resolution and long-range migration are up to $\sim 1\%$ for
all quantities. In the majority of cases the errors will be actually
much smaller.

\section{Multi-planet simulations}
\label{sec:multiplanet}
As pointed out in the introduction, part of the motivation for
developing this method is to capture the large temporal and spatial
ranges involved in systems with multiple planets.

In this multi-planet case, the borders of the mesh are computed
similarly as was done in the single-planet case. We obtain the
semi-major axis for all of them and then we calculate $r_d^-$ and
$r_b^-$ with Eqs.~\eqref{eq:damping} and \eqref{eq:border1} using the
semi-major axis of the innermost planet, while the same procedure is
followed for $r_d^+$ and $r_b^+$, using the outermost planet.

\begin{figure*}
	\centering
	\includegraphics[width=\textwidth]{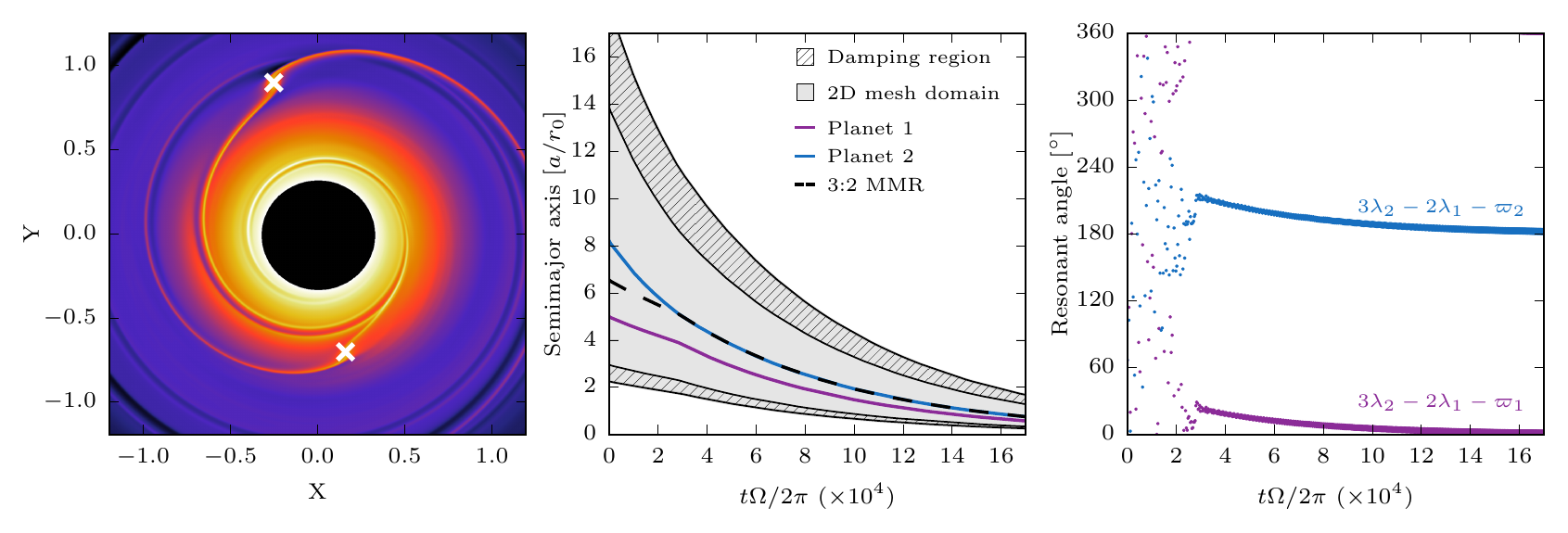}
	\caption{\label{fig:fig_2p} Hydrodynamical simulation for a
          two-planet system with masses $m_1 = 10M_{\oplus}$ and $m_2
          = 20M_{\oplus}$. Initial semi-major axes are $a_1 = 5$~AU
          and $a_2 = 8.19$~AU, and initial eccentricities equal to
          $0.05$. Disk parameters are $\Sigma_0 = 400$ gr/cm$^2$,
          $f=0$, $\alpha = -0.5$, $h = 0.05$ and $\alpha_{\nu} =
          4\times 10^{-3}$. The left frame shows the disk's surface
          density at $t = 15.53\times 10^4$ orbits, where the
          planets are indicated with white crosses. The middle frame
          shows the time evolution of the semi-major axes. The light
          gray region corresponds to the active mesh and the hatched
          region to the damping zones. Approximately at $t =
          3\times10^4$ orbits, both planets are trapped in 3:2 MMR,
          after which both resonant angles begin to librate around
          fixed values that slowly evolve with time (right plot).  }
\end{figure*}

\begin{figure*}
	\centering
	\includegraphics[width=\textwidth]{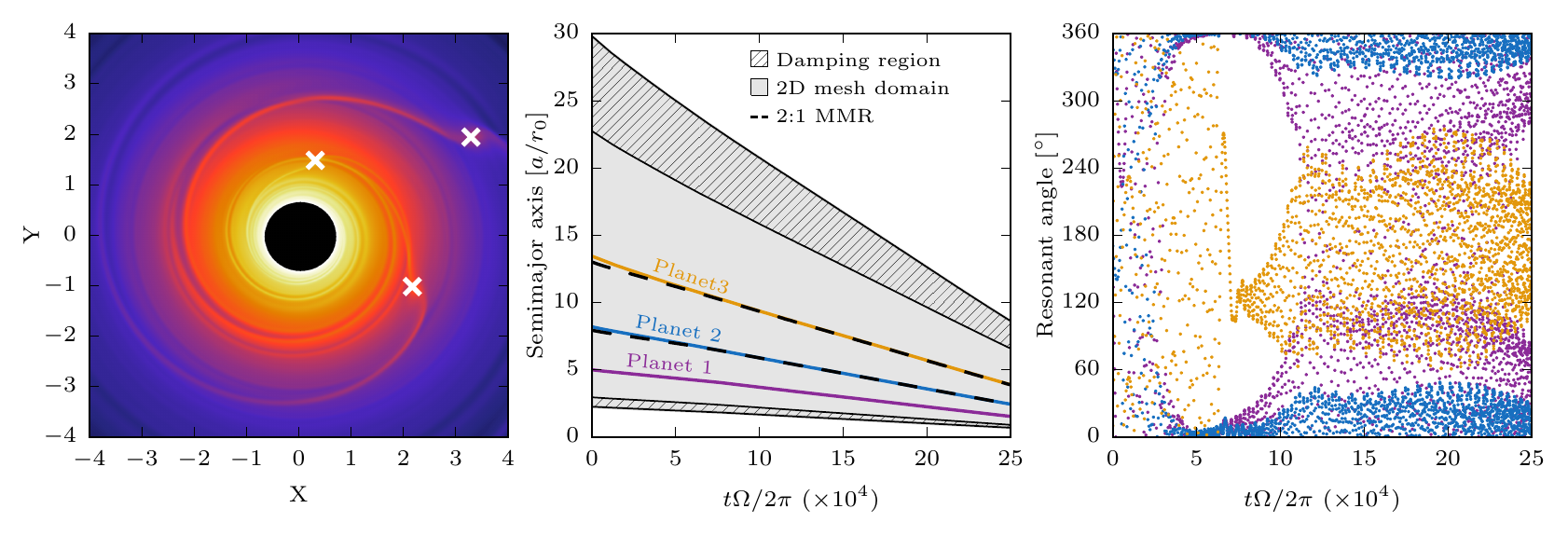}
	\caption{\label{fig:fig_3p} Hydrodynamical simulation of a
          three-planet system with masses $m_1 = 10M_{\oplus}$, $m_2 =
          20M_{\oplus}$ and $m_3 = 40M_{\oplus}$. Initial semi-major
          axes are $a_1 = 5$~AU, $a_2 = 8.19$~AU and $a_3 =
          13.44$~AU, and initial eccentricities are equal to
          $0.05$. Disk parameters are $\Sigma_0 = 400$ gr/cm$^2$,
          $f=0.25$, $\alpha = -1.0$, $h = 0.05$ and $\alpha_{\nu} =
          4\times 10^{-3}$.  The left frame shows the disk's surface
          density at $t = 25\times 10^4$ orbits, and the planets
          indicated with white crosses. The middle frame shows the
          time evolution of the semi-major axes.  The light gray
          region corresponds to the active mesh and the hatched region
          to the damping zones.  All planets are captured in
          successive 2:1 MMRs leading to a multiple-resonant
          configuration that it preserved throughout the
          simulation. The right plot shows the behavior of the main
          resonant angle of the inner pair $2\lambda_2 - \lambda_1 -
          \varpi_1$ (purple), the corresponding angle of the outer
          pair $2\lambda_3 - \lambda_2 - \varpi_2$ (blue), and the
          Laplace resonant angle $\lambda_1 -3\lambda_2 + 2\lambda_3$
          in orange.}
\end{figure*}

Fig.~\ref{fig:fig_2p} and Fig.~\ref{fig:fig_3p} show two examples, the
first corresponding to a two-planet system, while the second explores
a case with three planets. Initial conditions for the bodies and the
disk are detailed in the captions. In both cases the dynamical
evolution is carried out for long timescales such that the final
semi-major axes lie very close to the central star. The systems
naturally evolve to resonant configurations, as shown by the behavior
of the resonant angles.

\section{Conclusions}
\label{sec:conclusions}

We have presented a new method that allows long-term and large-scale
hydrodynamical simulations of migrating planets over a grid-based
Eulerian code.  Our method consists in the following steps:
\begin{enumerate}
\item A prescription for calculating the new radial borders of the
  mesh (see \S \ref{subsec:borders})
\item An interpolation method for filling the new mesh (see \S
  \ref{subsec:filling}).
\item Damping boundaries to avoid reflections and match the
  large-scale profiles (Eq. \ref{eq:damping}).
\item A method to un-shift the Lindblad resonances (see \S
  \ref{subsec:unshift}).
\item Optionally, a method to follow the global viscous evolution of
  the disk (see \S \ref{subsec:viscous}).
\end{enumerate}
We have validated our method with a set of tests using the public code
FARGO3D and compared its results with those obtained using classical
calculations with a fixed grid.

The remapping technique allows to achieve cleaner results compared to
the standard case at lower computational cost. Due to the adaptive
property of the algorithm described to compute the border of the mesh,
the resolution naturally adapts when the planets migrate, allowing for
a self-similar behavior of the system.

Our method allows for the first time to perform simulations not
limited by the radial extent of the mesh. This feature enables the
possibility of performing long-range and long-term hydrodynamical
simulations of planetary systems, during a time comparable to the
lifetime of the disk.  It is important to underline that our algorithm
lies on the hypothesis that we can derive the structure for the disk
outside the active mesh to fill the damping zones and ghost cells.

We have also discussed how to implement the viscous evolution of the
disk, which is important when simulations are long-term. However,
since such simulations are also usually long-range, a realistic model
for the disk should also be considered with a realistic non-power law
for surface density and the temperature.

We have discussed a possible improvement to the method, such as a
logarithmic interpolation to reduce errors. However, this
implementation is not strictly necessary and, in the current state,
the method has a largely sufficient accuracy for practical purposes.

\begin{acknowledgements}
  PBLL, XR and CB acknowledge financial support from CONICET and
  SECYT/UNC as well as computational resources provided by IATE and
  CCAD (Universidad Nacional de C\'ordoba). FM acknowledges support
  from CONACyT grant 178377. We thank Ulises Amaya Olvera, Reyes
  Garc\'\i a Carre\'on, and J\'er\^ome Verleyen for their assistance
  in setting up the GPU cluster on which many of the calculations
  presented here have been run.  We thank the referee for a very
  helpful report. His/her suggestions allowed us to significantly
  improve this work.
\end{acknowledgements}

\end{document}